\documentclass[twocolumn]{revtex4}
\usepackage{graphicx}

\begin{document}

\title{Sub-threshold channels at the edges of nano-scale triple-gate silicon transistors}

\author{H. Sellier}
\altaffiliation[Present address: ]{Laboratoire de Spectrom\'etrie
Physique, UJF-CNRS Grenoble, France. Electronic mail:
hermann.sellier@ujf-grenoble.fr}
\author{G. P. Lansbergen}
\author{J. Caro}
\author{S. Rogge}
\affiliation{Kavli Institute of Nanoscience, Delft University of
Technology, Lorentzweg 1, 2628 CJ Delft, The Netherlands}

\author{N. Collaert}
\author{I. Ferain}
\author{M. Jurczak}
\author{S. Biesemans}
\affiliation{InterUniversity Microelectronics Center (IMEC),
Kapeldreef 75, 3001 Leuven, Belgium}


\begin{abstract}

We investigate by low-temperature transport experiments the
sub-threshold behavior of triple-gate silicon field-effect
transistors. These three-dimensional nano-scale devices consist of
a lithographically defined silicon nanowire surrounded by a gate
with an active region as small as a few tens of nanometers, down
to $50\times60\times35$\,nm$^3$. Conductance versus gate voltage
show Coulomb-blockade oscillations with a large charging energy
due to the formation of a small potential well below the gate.
According to dependencies on device geometry and thermionic
current analysis, we conclude that sub-threshold channels, a few
nanometers wide, appear at the nanowire edges, hence providing an
experimental evidence for the corner-effect.

\end{abstract}


\maketitle


Non-planar field effect transistors called
FinFETs~\cite{Hisamoto:00} are currently being developed to solve
the problematic issues encountered with the standard planar
geometry when the channel length is reduced to a sub-100\,nm size.
Their triple-gate geometry is expected to have a more efficient
gate action and to solve the leakage problem through the body of
the transistor, one of the dramatic short channel
effects~\cite{Lemme:04}. However, their truly 3-dimensional (3D)
structure makes doping---and thus also potential---profiles very
difficult to simulate and to understand using the current
knowledge on device technology. Transport studies at low
temperature, where the thermally activated transport is
suppressed, can bring insight to these questions by measuring
local gate action. For this reason we experimentally investigate
the potential profile by conductance measurements and observe the
formation of a sub-threshold channel at the edge of the silicon
nanowire. This corner effect has been
proposed~\cite{Doyle:03,Fossum:03} as an additional contribution
to the sub-threshold current in these 3D triple-gate structures
where the edges of the nanowire experience stronger gate action
due geometric enhancement of the field. However, besides extensive
simulation work~\cite{Doyle:03,Fossum:03}---keeping in mind the
difficulties with these 3D structures---very little experimental
work has been published until now on this effect~\cite{Xiong:04}.


\begin{figure}[b]
\includegraphics[width=8cm,clip,trim=0 0 0 0]{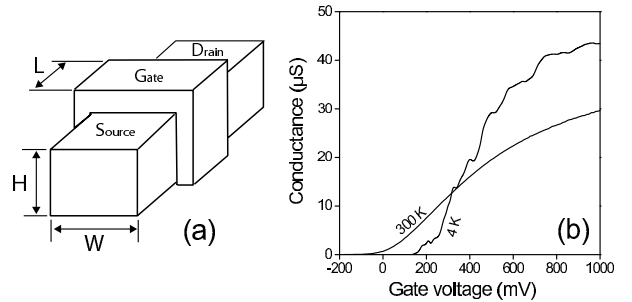}
\caption{(a) Schematic of the FinFET geometry where the gate
surrounds the Si nanowire (the fin). (b) Low bias conductance
versus gate voltage for a long and narrow silicon FinFET
($L=950$\,nm, $W=35$\,nm) at room and liquid helium temperatures.
Reproducible conductance fluctuations appear due to quantum
interferences.} \label{fig1}
\end{figure}


The FinFETs discussed here consist of a narrow single-crystalline
silicon wire with two large contact pads etched in a p-type
Silicon On Insulator (SOI) layer doped with $10^{18}$\,cm$^{-3}$
boron atoms. This silicon wire is covered with a
$t_\mathrm{ox}=1.4$\,nm thick thermal oxide and a second narrow
poly-crystalline silicon wire crossing the first one is fabricated
to form a gate that surrounds the wire on three faces (Fig.~1(a)).
The entire surface is then implanted with $10^{19}$\,cm$^{-3}$
arsenic atoms to form n-type degenerate source, drain, and gate.
During this implantation the wire located below the gate is
protected and remains p-type. In the investigated device series
the height of the fin wire is $H=60$\,nm, while the width ranges
from $W=35$\,nm to 1\,$\mu$m and the gate length ranges from
$L=50$\,nm to 1\,$\mu$m. The relatively high p-type doping of the
channel wire is chosen to ensure a depletion length shorter than
half the channel length in order to have a fully developed
potential barrier in this n-p-n structure and so to keep the
conductance threshold at a large enough positive gate voltage. The
characteristics at room temperature of these nano-scale FinFETs
look therefore similar to those of their larger planar
counterparts (Fig.~1(b) at 300\,K).


\begin{figure}[b]
\includegraphics[width=8cm,clip,trim=0 0 0 0]{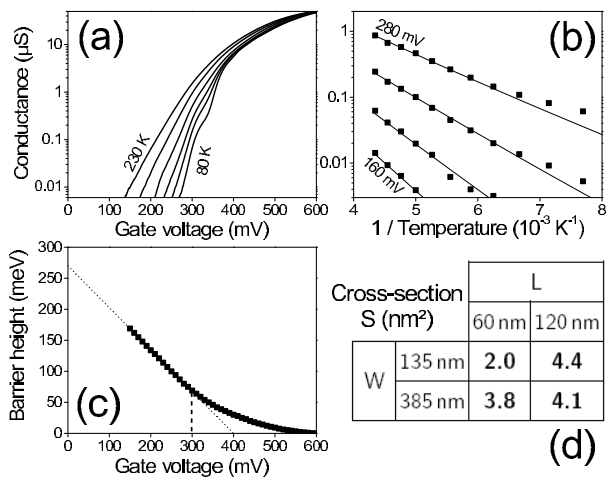}
\caption{(a) Conductance versus gate voltage (30\,K steps) for a
short and wide fin ($L=60$\,nm, $W=385$\,nm). (b) Conductance
plotted versus the inverse of the temperature (40\,mV steps) for
the same sample. The conductance is thermally activated above
150\,K. (c) Barrier height versus gate voltage changing behavior
at 300\,mV (same sample). (d) Measured cross-section $S$ for the
activated current of 4 samples with different length $L$ and width
$W$.} \label{fig2}
\end{figure}


For sub-threshold voltages, a barrier exists between the source
and drain electron reservoirs and the transport is thermally
activated at high enough temperature as shown in Fig.~2(b). For
very short devices, the conductance is simply given by the
thermionic emission above the barrier according to the formula:
$$ G=S\,A^*\,T\,(e/k_\mathrm{B})\exp{(-E_b/k_\mathrm{B}T)} $$
where the effective Richardson constant $A^*$ for Si is
$2.1\times120$\,Acm$^{-2}$K$^{-2}$~\cite{Sze:81}. Several samples
have been measured in this regime and their conductance has been
fitted to obtain the barrier height $E_b$ and the cross section
$S$ (Fig.~2(d)). The two 385\,nm wide samples have the same cross
section $S\approx4$\,nm$^2$ although their length differ by a
factor 2. We can therefore conclude that transport is dominated by
thermionic emission. The two 135\,nm wide samples however have
different $S$ values, but this can not imply a diffusive transport
since the longest sample has the largest conductance. Another
result is that the cross section $S\approx4$\,nm$^2$ is much
smaller than the channel width $W$ (135 or 385\,nm) multiplied by
the channel thickness (about 1\,nm). This result is consistent
with the corner-effect that produces a lower conduction band
(stronger electric field) along the two edges of the wire, where
the current will therefore flow preferentially (Fig.~3(b)). This
interpretation is confirmed by the result obtained on a 385\,nm
wide sample with an undoped channel. Its larger cross section
$S=24$\,nm$^2$ (still much smaller than the width) is in agreement
with the corner-effect since the longer depletion length of the
undoped silicon gives smoother potential variations and therefore
wider channels along the edges.


The barrier height $E_b$ versus gate voltage is plotted in
Fig.~2(c). The data extrapolated to zero gate voltage are
consistent with a 220\,meV barrier height calculated for a p-type
channel in contact with a n$^{++}$ gate through a 1.4\,nm SiO$_2$
dielectric~\cite{Sze:81}. The linear dependence of the barrier
height shows a good channel/gate coupling ratio $\alpha =
dE_b\,/\,e\,dV_G = 0.68$ due to the triple gate geometry with a
thin gate oxide. At higher gate voltage (above 300\,mV) the
coupling ratio decreases and a finite barrier survives up to large
voltages. Analysis of the low temperature transport (see below)
shows however that the gate action remains constant inside the
channel where localized states are formed. The finite barrier is
in fact two confining barriers located in the access regions
(between channel and contacts) where the concentration of
implanted arsenic atoms is reduced by the masking silicon nitride
spacers placed next to the gate (Fig.~3(a)).


\begin{figure}[b]
\includegraphics[width=8cm,clip,trim=0 0 0 0]{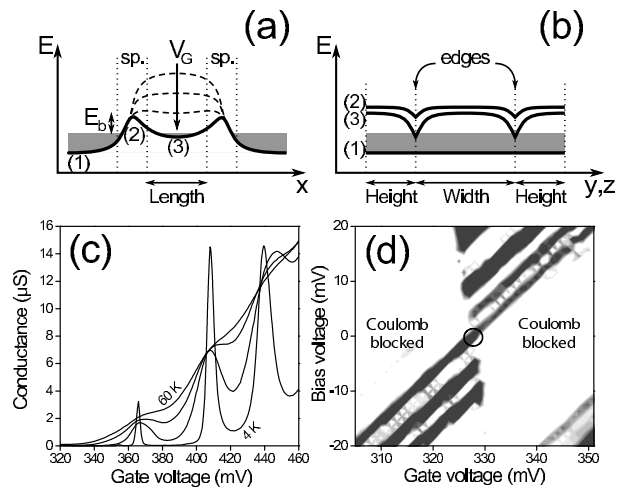}
\caption{(a) Conduction band edge profile with the highest barrier
in the channel or in the access regions below the spacers (sp.)
depending on gate voltage. (b) Band edge along the 3D gate oxide
interface (1) in the contacts, (2) in the barriers, (3) in the
channel. The corner-effect produces two channels with low barriers
at the wire edges. (c) Conductance versus gate voltage for a short
and wide channel ($L=60$\,nm, $W=385$\,nm) showing Coulomb
blockade peaks up to high temperatures (20\,K steps). (d)
Stability diagram, \textit{i.e.}~conductance versus gate and bias
voltages, at 4.2\,K. The circle indicates a zero bias conductance
peak, which develops into a triangular sector at finite bias.}
\label{fig3}
\end{figure}


At low temperature the conductance develops fluctuations versus
gate voltage (see Fig.~1(b)) with a pattern that reproduces after
thermal cycling (at least for the main features). These
fluctuations are caused by quantum interferences in the channel.
For gate voltages close to the threshold, charge localization
occurs, especially for short fins as can be seen in Fig.~3(c).
When cooled down to 4.2\,K the conductance pattern develops a
series of peaks that we attribute to Coulomb blockade of electrons
in the potential well created in the channel by the two tunnel
barriers of the low-doped access regions~\cite{Boeuf:03,Jehl:03}.
This interpretation is supported by the channel-length dependence
of the peak spacing discussed later. An explanation in terms of a
quantum well formed by an impurity can be ruled out. An impurity
or defect could not accept many electrons, e.g. more than 20 for
the 100\,nm sample in Fig.~4(b), since they represent a single
charge or empty state. These devices act therefore as quantum dots
where the conduction electrons are spatially localized and are
Coulomb blocked for the transport by a finite charging
energy~\cite{Kouwenhoven:93}.


In the stability diagram of a quantum dot (see Fig.~3(d)), the
slopes of a triangular conducting sector give the ratios of the
capacitances $C_G$, $C_S$, and $C_D$ between the dot and
respectively the gate, source, and drain
electrodes~\cite{Kouwenhoven:93}. In this way we find the dot/gate
coupling $\alpha=C_G/(C_G+C_S+C_D)=0.78$ (0.65) for the first
(second) resonance. These values are close to the channel/gate
coupling 0.68 obtained independently in the same sample from the
gate voltage dependence of the barrier height in the middle of the
channel at higher temperatures. This result indicates that the
gate coupling in the center of the device remains constant and
supports the idea of a minimum in the conduction band as sketched
in Fig.~3(a).


\begin{figure}[t]
\includegraphics[width=8cm,clip,trim=0 0 0 0]{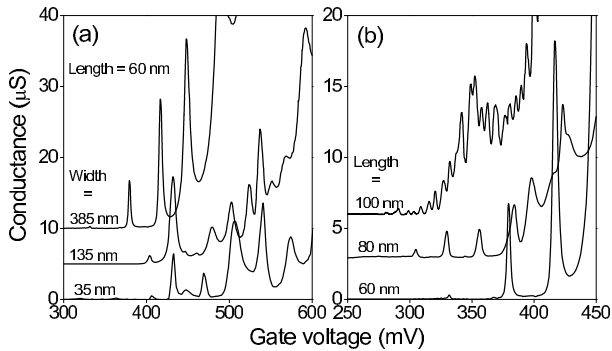}
\caption{Conductance versus gate voltage at 4.2\,K for several
devices. (a) Short fins ($L$ = 60\,nm) of different widths ($W$ =
35, 135, 385\,nm) have a similar peak spacing. (b) Devices with
longer fins ($L$ = 60, 80, 100\,nm) have a smaller peak spacing
(the widths are different). The curves have been shifted for
clarity.} \label{fig4}
\end{figure}


The peak spacing $\Delta V_G$ is the change in gate voltage that
increases by one the number of electrons in the dot located at the
silicon/oxide interface. This quantity provides the dot/gate
capacitance $C_G=e/\Delta V_G$, and then the dot area
$S=C_G/C_{ox}$ using the gate capacitance per unit area
$C_{ox}=\epsilon_{ox}/t_{ox}=0.025$\,F/m$^2$. The peak spacings
for the same gate length ($L$ = 60\,nm) but three different
channel widths ($W$ = 35, 135, and 385\,nm) can be compared in
Fig.~4(a). Although the patterns are not very regular, an average
peak spacing of about 30\,mV is obtained for all of them,
indicating similar dot areas whereas the effective width is varied
by more than a factor three. On the opposite, the conductance
patterns for three different lengths ($L$ = 60, 80, and 100\,nm)
shown in Fig.~4(b) have decreasing average peak spacings ($\Delta
V_G$ = 39, 24, and 6\,mV respectively) and therefore increasing
dot areas ($S$ = 160, 270, and 1100\,nm$^2$). However these areas
are not strictly proportional to the gate length, so that the
actual width could be length dependent or the actual dot length
could be smaller than the gate length for very short fins. If we
assume that the dot length equals the gate length, we obtain 2.7,
3.4, and 11\,nm for the dot width, \textit{i.e.}~a small fraction
of the total Si/oxide interface width $W_{eff}=W+2H$ = 150 to
500\,nm. The observation of similar dot widths of a few nm for
different fin widths of hundreds of nm is consistent with the idea
of a dot located at the edge of the fin and thus with the corner
effect~\cite{Doyle:03,Fossum:03}.


In addition to a large charging energy $E_c=\alpha\,e\,\Delta
V_G$, these dots also have a large quantum level spacing $\Delta
E$ as can be deduced from the temperature dependence of the
conductance peaks in Fig.~3(c). When the temperature is lowered
below the level spacing, the tunneling process involves a single
quantum level at a time and the peak height starts to increase
above the high temperature value~\cite{Beenakker:91}. The
crossover from the classical to the quantum regime of Coulomb
blockade being around 15\,K, we estimate the level spacing to be
about 1.3\,meV. If we use the gate length $L=60$\,nm in the
expression $\Delta E=3\,\pi^2\hbar^2/2m^*L^2$ for the energy
separation between the first and second states of a 1D system, we
find a level spacing $\Delta E=1.6$\,meV similar to the
experimental estimation. This result supports the idea of a long
dot extending over the whole gate length (assumed above to extract
the dot width from the dot/gate capacitance).


In conclusion, both the activated current amplitude, the Coulomb
blockade peak spacing, and the quantum level spacing reveal that
the current flows in narrow channels a few nanometers wide. They
appear along the edges of the FinFET due to an enhanced
band-bending called corner-effect. In order to get an homogeneous
current distribution with a lower sub-threshold current and a
larger on/off current ratio, this effect should be reduced. Better
devices would have rounder corners on the scale of the depletion
length and a lower doping concentration in the channel.


\end{document}